\documentclass[twocolumn,showpacs,prb]{revtex4}
\usepackage{graphicx}
\usepackage{amsmath}

\begin{document}

\title{
Approximate and exact nodes of fermionic wavefunctions:
coordinate transformations and topologies
}

\author{Michal Bajdich, Lubos Mitas, Gabriel Drobn\'y, and Lucas K.
Wagner}
\affiliation{ Center for High Performance Simulation and 
Department of Physics,
North Carolina State University, Raleigh, NC 27695.}

\date{\today}

\begin{abstract}
A study of fermion nodes  
for spin-polarized states of a few-electron
ions and molecules with $s,p,d$ one-particle
orbitals is presented. We find exact nodes for some cases of 
two electron atomic and 
molecular states and also
the first exact node for the three-electron
atomic system in  $^4S(p^3)$ state using appropriate coordinate
maps and wavefunction symmetries.   
We analyze the cases of nodes for larger number
of electrons in the Hartree-Fock approximation
and for some cases we find transformations for
projecting the high-dimensional node manifolds into
3D space. The node topologies and other properties are studied
using these projections.  We also propose a general coordinate 
transformation as an extension of Feynman-Cohen
backflow coordinates to both simplify the nodal description and
as a new variational freedom for quantum Monte Carlo
trial wavefunctions.
\end{abstract}


\maketitle

\section{\label{sec:level1} Introduction}
The problem of fermion nodes
is one of the most intriguing
challenges in quantum simulations
of fermionic systems by stochastic methods such as
quantum Monte Carlo (QMC)\cite{qmchistory,hammond}. 
In QMC the many-fermion wavefunction is represented by an ensemble of sampling
points (walkers) in the space of fermion coordinates. 
The walkers are propagated
 according to the matrix elements of the projector
$\exp(-\tau H)$ where $\tau$ is a real parameter
and $H$ is a Hamiltonian. It is straightforward
to show that the propagated wavefunction solves
the time-dependent
Schr\"odinger equation in imaginary time $\tau=it$
and converges to the ground state for $\tau\to \infty$.
Unfortunately, for fermions such a straightforward
sampling process
runs into difficulties and the projection 
becomes very inefficient. 
The ensembles of walkers which
initially
sample negative and positive parts of the wavefunction
are independent and asymptotically converge
to the same (bosonic) distribution with
an exponential growth of error bars
for fermionic observables.
In electronic structure QMC 
calculations\cite{qmcrev}
this well-known fermion sign problem is
circumvented
by the fixed-node approximation which restricts the negative and
positive walkers into separate regions of space defined by an approximate
fermion
node (zero boundary) of the best available trial/variational wavefunction.
That guarantees stability of the statistical error bars
at the price of a fixed-node bias. The fixed-node bias is
proportional to the square of  
 the nodal displacement error and therefore in typical
electronic structure calculations the resulting bias is 
 rather small. 
Even for hundreds of electrons, 
Hartree-Fock or multi-reference Hartree-Fock (HF)
nodes lead to impressive accuracy when used within QMC. The 
 fixed-node QMC calculations
typically  provide about 95\% of the correlation energy in real
systems \cite{qmcrev} such as molecules, clusters, solids, etc.

At a more fundamental level, knowledge of the exact node
enables one to eliminate the fixed-node bias completely, and 
the exact energy can be calculated in time which scales 
as a low-order polynomial in the number of particles.
Therefore elimination of the fixed-node error remains one
of the intriguing possibilities  for employing 
QMC to attack
a number of important many-body problems which require accuracy beyond
a few-determinant Hartree-Fock nodes.

Let us assume 
 a system of spin-polarized electrons  
described by a real wavefunction $\Psi(R)$ where 
 $R$ denotes the electron spatial coordinates. The exchange of 
an electron pair with labels $i,j$, denoted as $P_{ij}$, 
gives $\Psi(R) = - \Psi(P_{ij}R)$. Consequently, the
antisymmetry implies that  
there exists a subset of electron configurations, 
called a fermion node,
for which the wavefunction is zero. Let us eliminate
the regions in which the wavefunction vanishes because of other
reasons (eg, external potential);  then the 
fermion node is given by an implicit
equation $\Psi(R)=0$.  In general, the fermion 
node is 
a ($ND$-1)-dimensional manifold (hypersurface) assuming that we have
$N$ fermions in a $D-$dimensional space.
The fermion nodes of small systems, mostly atoms, were investigated in several
previously published  papers 
\cite{node1s2s,dario04,jbanderson75,andersp,lesternode}.
The general properties
of fermion nodes were analyzed in an extensive study by 
Ceperley \cite{davidnode} which included a proof of the tiling 
property and generalizations of the fermion nodes to density matrices.
We mention two of the results which will be used later.

i) Tiling property for the nondegenerate ground state: Let us
define a nodal cell
 $\Omega(R_0)$ as a subset of configurations which can be reached
from the point $R_0$ by a continuous path without crossing the node. 
The tiling property 
says that by applying all possible particle permutations to an arbitrary 
nodal cell
of a ground state wavefunction one covers the complete configuration
space. Note that this does not specify how many nodal cells are there.
 Furthermore, symmetry of the state is also
symmetry of the node and tiling property is valid for any non-degenerate 
ground state
within the given discrete symmetry.

ii) If two nodal surfaces cross each other they are orthogonal
at the crossing. If $n$ nodal surfaces cross each other, the crossing 
angles are all equal
 to $\pi/n$. 

In addition,  it was numerically shown \cite{davidnode}
that the nodal cells for a number of finite
 free particle systems are maximal,
ie, all regions with the same sign of wavefunction are interconnected.  
The fermion nodes for degenerate and excited states were further
studied by Foulkes and co-workers \cite{foulkes}. 
Recent interesting work by Bressanini, 
Reynolds and Ceperley revealed differences in the nodal surface topology
between Hartree-Fock and correlated wavefunctions for the Be atom 
explaining the large impact of the $2s,2p$ near-degeneracy 
on the fixed-node QMC energy \cite{dariobe}.

The remainder of this paper is organized as follows: in Sec. \ref{sec:level2}
we discover new {\em exact} fermion nodes for two and three-electron spin polarized systems.
In Sec. \ref{sec:level3} we categorize the nodal surfaces for the several half-filled 
subshells relevant for atomic and molecular states. In Sec. \ref{sec:level4} 
we suggest a general particle position transformation, 
both as a tool to simplify the description of the nodes and also
as a possible new variational freedom for trial wavefunctions. Finally, 
in the last section we present our conclusions 
and suggestions for future work.

\section{\label{sec:level2} Exact nodal surfaces}
We assume the usual electron-ion Hamiltonian and we first 
investigate a few-electron ions focusing on fermion nodes for
subshells of one-particle states with $s,p,d,f ... $ symmetries using 
variable transformations, symmetry operations and explicit
expressions for the nodes.

\subsection{\label{sec:level21}Three-electron quartet $ ^4S (p^3)$ 
state}
\begin{figure}[ht]
\centering
{\resizebox{3.4in}{!}{\includegraphics{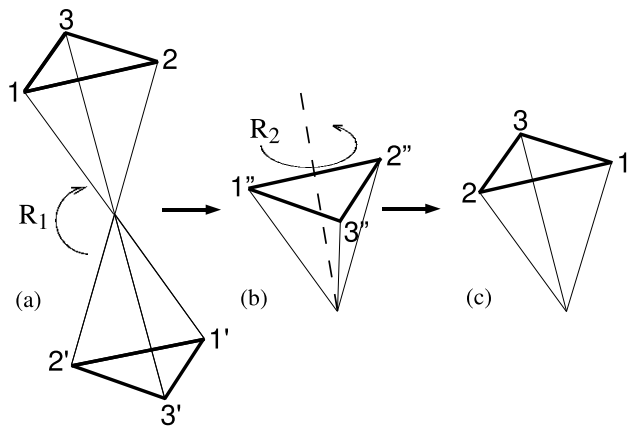}}}
\caption{Inversion and two subsequent rotations of three particles:
(a) Original and inverted (primed) positions;
(b) Positions after the rotation $R_1$  
in the plane given by the particles 1,2 and the origin;
(c) Positions after the second rotation $R_2$
around the ${\bf r}_{1}+{\bf r}_2$ axis. Note that
the original positions of
the  particles 1 and 2 are exchanged. 
\label{fig:0}}
\end{figure}

Let us first analyze a special case with $r_1=r_2$ and 
$r_{23}=r_{31}$. It is then easy to see that the 
inversion around the origin with subsequent rotations is equivalent
to the exchange of two particles, say, 1 and 2 (Fig.\ref{fig:0}). 
Therefore for this particular configurations of particles
the combination of parity and rotations 
is closely related to the exchange symmetry.
The illustration also shows that the six distances
do not specify the relative positions of the three electrons
unambiguously. For a given set of the distances there 
are two distinct positions, say, of the electron 
3, relative to the fixed positions
of electrons 1 and 2  (see Fig.\ref{fig:0}) and compare positions
3 and 3'' of the third electron.

\begin{figure}[ht]
\centering
{\resizebox{3.4in}{!}{\includegraphics{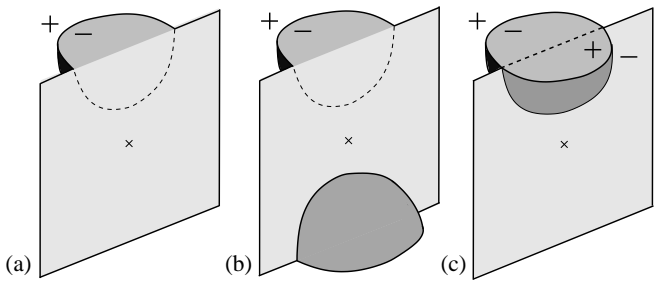}}}
\caption{ (a) An illustration of an artificial
  distortion of the planar ground state node
for the quartet state (see text);
(b) The original and parity transformed distorted node; 
(c) Finally, a subsequent 
rotation of the inverted distortion necessarily leads to  
a new nodal pocket which is artificial for the ground state. 
In fact, nodes with similar topologies are present
in excited states \cite{lubosmc28}.
\label{fig:blob}}
\end{figure}

In order to analyze the wavefunction in an unambiguous manner 
it is convenient to define new coordinates.
Let us denote
${\bf r}_{12}^+ ={\bf r}_1+{\bf r}_2, r_{12}^+=|{\bf r}_{12}^+| $,
together with the customary 
${\bf r}_{12} ={\bf r}_1-{\bf r}_2, r_{12}=|{\bf r}_{12}| $ . 
We can now introduce the following 
map of the Cartesian coordinates 
\begin{equation}
({\bf r}_1, {\bf r}_2, {\bf r}_3) \to
(r_{12}^+, r_{12}, r_3, \cos\alpha,\cos\beta,\gamma,\Omega)
\end{equation}
with definitions: $\cos\alpha=
{\bf r}_3\cdot({\bf r}_1\times {\bf r}_2)/(r_3|{\bf r}_1\times {\bf r}_2|)$,
$\cos\beta={\bf r}_{12}^+\cdot {\bf r}_{12}/(r_{12}^+ r_{12})$ and
$\gamma$ being an azimuthal angle of ${\bf r}_3$ in the
relative coordinate system with unit vectors
${\bf e}_x= {\bf r}_{12}^+/r_{12}^+$, 
${\bf e}_z={\bf r}_1\times {\bf r}_2/|{\bf r}_1\times {\bf r}_2|$, 
${\bf e}_y={\bf e}_z\times {\bf e}_x$.
For completeness, 
$\Omega$ denotes three Euler angles which fix the orientation the 
three-particle system in the original coordinates  
(eg, two spherical angles of ${\bf r}_1\times {\bf r}_2$
and an azimuthal angle of ${\bf r}_{12}^+$). Since the angles $\Omega$
are irrelevant in $S$ symmetry, the
 first six variables fully specify the relative positions
of the three particles and
the wavefunction dependence simplifies to
$\Psi(r_{12}^+, r_{12}, r_3, \cos\alpha,\cos\beta,\gamma)$.
Consider now two symmetry operations which change the sign of the wavefunction and
keep the distances unchanged: parity $P_I$
and exchange $P_{12}$ between particles 1 and 2. 
The exchange flips the sign of all three $\cos\alpha,\cos\beta,\gamma$ 
while the parity changes only the sign of $\cos\alpha$.
The action of $P_I P_{12}$ on $\Psi$ leads to
\begin{equation}
\Psi(...,\cos\alpha,-\cos\beta,-\gamma)=
\Psi(...,\cos\alpha,\cos\beta,\gamma)
\end{equation}
showing that the wavefunction is even in the simultaneous sign flip
$(\cos\beta,\gamma) \to (-\cos\beta,-\gamma)$.
Applying the exchange operator $P_{12}$ to the wavefunction 
and taking advantage of the previous property gives us
\begin{equation}
\Psi(...,-\cos\alpha,\cos\beta,\gamma)=
-\Psi(...,\cos\alpha,\cos\beta,\gamma)
\end{equation}
suggesting that there is a node determined by the condition $\cos\alpha=0$. 
It is also clear that the same arguments can be repeated
with
exchanged particle labels $2\leftrightarrow 3$ and $3\leftrightarrow 1$
and we end up with the {\em  the same nodal condition}:
${\bf r}_3\cdot({\bf r}_1\times {\bf r}_2)=0$. This shows that
 the node is encountered
when all three electrons lie on a plane passing
through the origin. 
Now we need to prove that this is the only node 
since there might possibly be
other nodal surfaces not revealed by the parametrization above.
The node given above clearly fulfills the tiling 
property and all symmetries of the state. 
Furthermore, the state is the lowest quartet of 
$S$ symmetry and odd parity (lower quartets such as $1s2s3s$,$1s2s2p$,
and $1s2p^2$ have either different parity or symmetry) and for the
ground state
we expect that the number of nodal cells will be minimal.
This is indeed true 
since the node above specifies only two nodal cells
(one positive, one negative): an electron is either on one or 
the other side
of the nodal plane passing through the remaining two electrons.
Futhermore, any distortion of the node from the plane necessarily leads 
to additional nodal cells (see Fig. \ref{fig:blob}) which 
can only increase energy by imposing higher curvature 
(kinetic energy) on the wavefunction.
This is basically the Feynman's argument from the proof 
demonstrating that the energy
of fermionic ground state is always above the energy of the bosonic 
ground state
(and
also essentially the same 
argument as used for the proof of the tiling property\cite{davidnode}).
In fact, as we show in another paper,
higher excited states
of this symmetry have additional nodes, as expected
\cite{lubosmc28}.
Given all the arguments above we conclude the proof that the plane
is the {\em exact node}.
Note that it is identical to the node of Hartree-Fock 
wavefunction of $2p$ orbitals given by  
$\Psi_{HF}={\rm det} \{\rho(r)x, 
\rho(r)y, \rho(r)z\}$ where $\rho(r)$ is a radial function.

The coordinate transformation above is not the only one which can 
be used to analyze this state. 
The high symmetry of the problem
enables us to find an alternative coordinate map 
with definitions of $\cos\beta$ modified to 
$\cos\beta'=[({\bf r}_1\times {\bf r}_2)\times {\bf r}_{12}^+]
\cdot  {\bf r}_{12}/
(|({\bf r}_1\times {\bf r}_2)\times {\bf r}_{12}^+|
|{\bf r}_{12}^+|)$
 and $\gamma$ to
$\gamma'$ by  redefinition of ${\bf e}_z$ to
${\bf e'}_z=[({\bf r}_1\times {\bf r}_2)\times {\bf r}_{12}]\times {\bf r}_{12}^+]
/|({\bf r}_1\times {\bf r}_2)\times {\bf r}_{12}]\times {\bf r}_{12}^+]|$
and ${\bf e'}_y={\bf e'}_z\cdot {\bf e}_x$.
In the redefined coordinates the search for the node
simplifies to an action of $P_{12}$
on $\Psi(r_{12}^+, r_{12}, r_3, \cos\alpha,\cos\beta',\gamma')$ 
\begin{equation}
\Psi(...,-\cos\alpha,\cos\beta',\gamma')=
-\Psi(...,\cos\alpha,\cos\beta',\gamma')
\end{equation}
since the distances and $\cos\beta', \gamma'$ are invariant to $P_{12}$.
Obviously, this leads to the same 
nodal condition as derived above.

It is quite interesting to compare these two coordinate maps
with $\beta,\gamma$ and $\beta', \gamma'$.
Although parity and exchange are independent operators the analysis above shows 
that in an appropriate coordinate system they imply the same nodal surface.
Both these operators cause an identical sign change
of the wavefunction indicating thus a special symmetry of the
$^4S(p^3)$ ground state node which is higher than would be expected
solely from antisymmetry. Similar observation was made in a study 
of fermion node in another case of two electron atomic state
\cite{andersp,darionew}.

\subsection{\label{sec:level22}Two-electron triplet $ ^3P (p^2)$  and  
$^3\Sigma_g(\pi^2)$ states} 
Apparently, the exact node of this case was derived in a different 
context by Breit in 1930 \cite{dario04,darionew}. 
Here we offer an independent
proof which enables us to apply  the analysis 
to some molecular states with the same symmetries.
The exact node for the $ ^3P (p^2)$ state can be found 
in a similar way as in the case of quartet above.
The state has even parity, 
cylindric symmetry, say,  around $z$-axis, and is odd
under rotation by $\pi$ around $x,y$ axes, 
$R(\pi x)$, $R(\pi y)$.
The mapping of  
Cartesian coordinates
which enables to analyze the wavefunction 
 symmetries is given by  
\begin{equation}
({\bf r}_1, {\bf r}_2) \to
(r_{12}^+,r_{12},\cos\omega,\cos\beta,\varphi,\varphi')
\end{equation}
where 
$\cos\omega=
{\bf z}_0\cdot({\bf r}_1\times {\bf r}_2)/
|{\bf r}_1\times {\bf r}_2|$
with
${\bf z}_0$ being the unit vector in the $z$-direction
and
$\varphi'$ being the azimuthal angle of ${\bf r}_{1}\times  {\bf r}_{2}$;
 $\varphi'$ can be omitted due to the cylindric symmetry.
Further, $\varphi$  is the azimuthal angle of ${\bf r}^+_{12}$ 
in the relative coordinate system with the $x$-axis unit vector
given by a projection of ${\bf z}_{0}$ into the plane
defined by ${\bf r}_1, {\bf r}_2$,
ie, ${\bf e}_x= {\bf z}_{0p}/|{\bf z}_{0p}|$, 
 ${\bf e}_z=({\bf r}_1\times {\bf r}_2)/|{\bf r}_1\times {\bf r}_2|$ and 
 ${\bf e}_y={\bf e}_z\times {\bf e}_x$.
Action of $P_IP_{12}R(\pi x)$ reveals that the wavefunction is
invariant in the simultaneous change  $(\cos\beta,\varphi)$ $\to$ 
$(- \cos\beta, -\varphi)$. This property and action of  
$P_{12}$ to the wavefunction together lead to 
\begin{equation}
\Psi(...,-\cos\omega,...)=
-\Psi(...,\cos\omega,...)
\end{equation}
with the rest of the variables unchanged.
The node is therefore given by $\cos\omega=0$
and is encountered when an electron hits the plane which contains the
$z$-axis and the other electron.
As in the previous case
 the nodal plane fulfills the tilling property and manifestly divides
the space into two nodal cells  so that we can conclude that this
node is exact.
The exact node again agrees with the node of Hartree-Fock wavefunction 
$\Psi={\rm det}\{\rho(r)x,\rho(r)y\}$. 

\begin{table}[ht]
\vspace{0.05cm}
\caption{Total energies (a.u.) of N$^{+}$, N$^{++}$ and N$^{+++}$ ions with core electrons
eliminated by pseudopotentials. The energies are calculated
by variational
(VMC) and fixed-node diffusion (DMC) quantum Monte Carlo and Configuration
Interaction (CI) methods. The HF energies are given as a reference for estimation of
the correlation energies.}
\begin{ruledtabular}
\begin{tabular}{l c c c c}
State  & HF  & CI  & VMC  & DMC \\
\hline
 $^3P(p^2)$  & -5.58528 &  -5.59491  &    -5.59491(2)  &  -5.59496(3)     \\
 $^4S(p^3)$  & -7.24716 &  -7.27566  &    -7.27577(1)  &  -7.27583(2)     \\
 $^5S(sp^3)$  & -8.98570 & -9.02027  &    -9.01819(4)  &  -9.01962(5)     \\
\end{tabular}
\end{ruledtabular}
\label{tab_xx}
\end{table}

The fixed-node QMC energies for the 
$^4S (p^3)$  and $^3P(p^2)$ cases derived above were calculated for 
a nitrogen cation
 with valence electrons in these states.
The core electrons were eliminated
by pseudopotential \cite{lester}. The trial wavefunction was
of the commonly used form with single HF determinant times a Jastrow correlation 
factor \cite{qmcrev}.
Note that the pseudopotential nonlocal $s-$channel does not couple
to either odd parity $S$ state or even parity $P(p^2)$ state so that
that the nonlocal contribution to the energy vanishes exactly.

In order
to compare the fixed-node QMC calculations with an independent 
method
we have carried out also 
Configuration Interaction (CI) calculations 
with ccpV6Z basis \cite{dunning} (with up to three
$g$ basis functions) which
generates more than 100 virtual orbitals in total. In the CI method the wavefunction
is expanded in excited determinants and we have included
all single, double  and triple excitations.
Since the doubles and triples include two- and three-particle correlations
exactly, the accuracy of the CI results is limited only by the size
of the basis set. By comparison with other two- and three-electron
 CI calculations we estimate that the order of magnitude of
the basis set CI bias is  $\approx$ 
0.01 mH (miliHartree) for two electrons and  $\approx$ 0.1  mH
and for three electrons (despite the large number of virtuals 
the CI expansion converges relatively slowly\cite{kutz} in the maximum
angular momentum of the basis functions, in our case $l_{max}=4$).
The pseudopotentials we used were identical in both QMC and CI 
calculations.

The first two rows of Tab.\ref{tab_xx} show the total energies of
variational and fixed-node DMC calculations 
with the trial wavefunctions with HF nodes together with
results from the CI calculations.
For $^3P(p^2)$ the energies agree within a few hundredths of mH 
with the CI energy being slightly higher but within two standard
deviations from the fixed-node QMC result.  For  
$^4S (p^3)$ the CI energy is clearly above the fixed-node DMC by about 
0.17 mH as expected due to the limited basis set size. In order to 
illustrate the effect of the fixed-node approximation in the case
when the HF node is {\em not} exact we have also included calculations
for four electron state $^5S(sp^3)$ (for further discussion
of this Hartree-Fock node see part III. below). For this case, 
we estimate
that the CI energy 
is above the exact one by $\approx 0.3$ mH so that the fixed-node energy 
is significantly {\em higher} than both CI and exact energies.
Using these results we estimate that the  
fixed-node error is  $\approx $ 1 mH, ie, close to 3\% of the correlation energy. 

Since in the $p^2$ case we have assumed 
cylindric symmetry, the derived node equation is applicable to any
such potential, eg, equidistant homonuclear dimer, trimer, etc,
with one-particle orbitals 
$\pi_x,\pi_y$ which couple into the triplet state
$^3\Sigma_g(\pi_x\pi_y)$.

Note that the parametrization given above 
automatically provides also one of the very few known
exact nodes in atoms so far \cite{node1s2s}, ie,
the lowest triplet state of He $^3 S (1s2s)$.
The spherical symmetry makes 
 angles $\omega$ and $\varphi$  irrelevant and simplifies
the two-electron wavefunction dependence to distances $r_1,r_2, r_{12}$ 
or, alternatively, to
 $r_{12},r_{12}^+,\cos\beta$.
Applying $P_{12}$ to wavefunction 
$\Psi(r_{12},r_{12}^+,\cos\beta)$
leads to
\begin{equation}
-\Psi(r_{12},r_{12}^+,\cos\beta)=
\Psi(r_{12},r_{12}^+,-\cos\beta) 
\end{equation}
so that the node is given by the condition $\cos\beta=0$, ie, $r_1-r_2=0$.

In addition, the presented analysis sheds some light on the He $^3P(1s2p)$
state node which
was investigated before \cite{andersp} as having higher symmetry
than implied by the wavefunction symmetries.
 The symmetry operations reveal that the wavefunction
depends on $|\cos\omega|$ and that the node is related to the
simultaneous flips such as  $(\cos\beta,\varphi)$ $\to$ 
$(- \cos\beta, -\varphi)$ or angle shifts
$\varphi \to \varphi+\pi$. Since, however, two of the variables are involved,
the node
has a more complicated shape as the previous study illustrates \cite{andersp}.
In order to test the accuracy of the HF node we have carried out a
fixed-node diffusion Monte Carlo
calculation of the  He $^3P(1s2p)$ state.  The resulting total energy of
 -2.13320(4) a.u. is in an excellent agreement with the estimated exact
value of -2.13316\cite{Schwartz}
which shows that the HF node is very close to
the exact one\cite{james}.

\section{\label{sec:level3}  Approximate Hartree-Fock nodes}
It is quite instructive to investigate the nodes of half-filled 
subshells of one-particle states with higher angular momentum.

\begin{figure}[ht]
\centering
{\resizebox{3.4in}{!}{\includegraphics{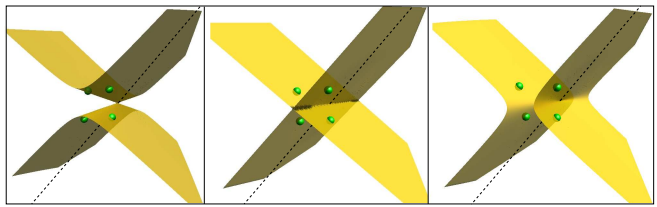}}}
\caption{(Color online) The 3D projected 
Hartree-Fock node of $^6S(d^5)$ state which is an elliptic cone
(left and right pictures). 
The middle picture illustrates a
case when two pairs of two electrons lie on orthogonal
planes which pass through the origin. This two-plane node is 
of lower dimension because of the additional condition 
on positions of the electrons. It appears
as a crossover between the cones with different
orientation (left 
and right pictures). The small spheres show the positions
of the four electrons while the line denotes the $z-$axis.
 \label{fig:d5}}
\end{figure}

\subsection{\label{sec:level31}Approximate Hartree-Fock node of $^6S(d^5)$ state}
The HF determinant wavefunction for $^6S(d^5)$
is given
\begin{equation}
\Psi_{HF}= \Pi_{i=1}^5 \rho(r_i) det\{2z^2-x^2-y^2,x^2-y^2,xz,yz,xy\}
\end{equation}
where $\rho(r_i)$ is the radial part of the $d$-orbital and 
we assume that all the orbitals
are from the same $l=2$ subshell, eg, $3d$ subshell. Since all radial functions
are the same they factor out from the determinant and for the 
purpose of finding the node they can be omitted. The
$S$ symmetry allows to rotate the system so that, say, electron 1 is
on the $z$-axis, and then the corresponding column in the Slater matrix
becomes $(2z_1^2,0,0,0,0)$. Assuming that $z_1\neq 0$ we 
can then write the nodal condition as
\begin{equation}
 det\{x^2-y^2,xz,yz,xy\} =0.
\end{equation}
Using one of the electrons as a 
{\em probe} (ie, looking at the node from the
perspective of one of electrons) we can find the projection 
of the node to 3D space.
By denoting the probe electron coordinates
simply as
$(x,y,z)$ and by expanding the determinant 
we get
\begin{equation}\label{eq:5dHFstart}
(x^2-y^2)m_1 +xzm_2+yzm_3+xym_4=0
\end{equation}
where 
$m_i$ are the corresponding cofactors. We divide out
the first cofactor assuming that it is nonzero 
(not a crucial assumption as clarified below).
We get 
\begin{equation}\label{eq:cone5d}
(x^2-y^2)+axz+byz+cxy=0
\end{equation}
where  $\; a=m_2/m_1$,  $b=m_3/m_1$, $c=m_4/m_1$.
By completing the square
this can be further rearranged to 
\begin{equation}\label{eq:cone5d2}
(x-k_1y)(x-k_2y) +z(ax+by) =0
\end{equation}
with
$k_{1,2}=(-c\pm\sqrt{c^2+4})/2$. 
Let us define
rotated and rescaled coordinates 
\begin{equation}
u^*=-(ak_2-b)(x-k_1y)/(k_1-k_2)
\end{equation}
\begin{equation}
v^*=(ak_1-b)(x-k_2y)/(k_1-k_2)
\end{equation}
\begin{equation}
w^*=z[(ak_1-b)(ak_2-b)]/(k_1-k_2)^2
\end{equation}
so we can write the Eq. (\ref{eq:cone5d}) as 
\begin{equation}\label{eq:5dHFgeneral}
u^*v^* +w^*u^* +w^*v^*=0.
\end{equation}
Note that this equation has a form which is
 identical to Eq. (\ref{eq:5dHFstart}) with 
$m_1=0$ so this representation is correct for general $m_1$.
After some effort one finds that Eq. (\ref{eq:5dHFgeneral}) is
a cone equation (ie, $d_{z^2}$ orbital) as can be easily
verified by using the following identity
\begin{equation}
(2u^2-v^2-w^2)/8=u^*v^* +w^*u^* +w^*v^*
\end{equation}
where  $ u=u^*+v^*+2w^*$, $ v=(-u^*+v^*+2w^*)$,
$w=(u^*-v^*+2w^*)$. 
 The 3D projected node is therefore rotated and rescaled 
 (elliptic) cone. 

At this point it is useful to
clarify how
the derived node projection cone
is related to the complete 14-dimensional node.
Remarkably, the 3D projection
 enables us to understand some of the properties
of the 14-dimensional manifold. 
First, the cone orientation 
and elliptic radii (ie, rescaling of the 
two axes with respect
to the third one) are determined
by the position of the four electrons in 3D space: 
with the exception of special lower dimensional cases explained below
there always exists a unique cone given by the  Eq. (\ref{eq:5dHFgeneral})
which "fits" the positions of the four electrons. 
Besides the special cases (below)
we can therefore define
a projection of a single point
in $4\times 3=12$-dimensional space of four electrons  
onto a cone. That also implies that the complete
12-dimensional space describes a set (or family) of cones which
are 3D projections of the nodal manifold.
Similar projection strategies are often used in algebraic geometry
to classify or analyze surfaces with complicated
topologies and/or high dimensionalities.

Since the cone orientation and two radii
 are uniquely defined by the point in 12 dimensions
and the cone itself is a 2D 
 surface in 3D space of the probe electron the complete
node then has 12+2=14
dimensions.  Therefore the $d^5$ HF node
is a set of cone
surfaces specified by the positions of the electrons . 
This 
particular form
is simply a property of the 
$d^5$  Hartree-Fock determinant.
From the derivation above it is clear that after factoring 
out the radial parts one
obtains
a homogeneous second-order polynomial in
three variables with coefficients determined by the 
positions of the four
electrons. 
In fact, from the theory of quadratic surfaces 
\cite{rektorys}, one finds
that a general elliptic cone can possibly fit up to 
five 3D points/electrons, 
however, in our case the cone has an additional constraint.
Our system was reoriented so
that one of the electrons lies on the $z$-axis;  that 
implies that the $z$-axis lies on the cone.
Therefore the cone always 
cuts the $xy$  (ie, $z=0$) plane in two lines which are orthogonal to each other.
The orthogonality can be verified by imposing
$z=0$ in Eq. (\ref{eq:cone5d2}) and checking that $k_1k_2=-1$. 
In addition, one can find
 "degenerate" configurations with two pairs of two electrons
lying on orthogonal planes (Fig. \ref{fig:d5}).
This corresponds to the "opening" of the cone 
 with one of the elliptic radii becoming
infinite and the resulting node having a form of
 two orthogonal planes (Fig. \ref{fig:d5}).
Since in this case there is an additional condition 
on the particle positions, the two-plane node has
lower dimension and is a zero measure 
subnode of the general 14-dimensional node.
The condition is equivalent to
$A_{44}=b^2-a^2-abc=0$,  where
$A_{44}$ is one of the quadratic invariants \cite{rektorys}.
There are more special cases of lower dimensional nodes:
a) when two electrons lie on a straight line going through
the origin;
 b) when three electrons lie on a plane
going through the origin; c) when four electrons lie in a single plane.

Remarkably, the analysis above enables us to find the number
of nodal cells. From Fig. \ref{fig:d5} one can infer
that by appropriate repositioning of the four electrons 
the cone surface smoothly "unwraps" the domains inside the cone, forms
two crossing planes and then "wraps" around the cone domains of the opposite
sign. That implies that an electron inside one of the cone
regions
can get to the region outside of the cone (with the same 
wavefunction sign) without any node crossing, using only appropriate  
concerted repositioning of the remaining four electrons.
That enables us to understand that a point in the
15-dimensional space (positions of five electrons)
 can continuously scan 
the plus (or minus) domain of the wavefunction:
 there are only two maximal nodal cells.

\begin{figure}[ht]
\centering
{\resizebox{3.4in}{!}{\includegraphics{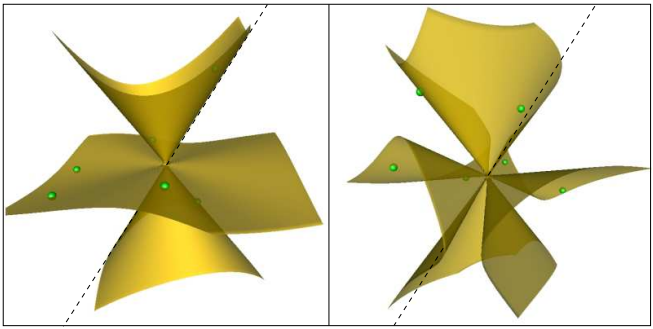}}}
\caption{(Color online) Projected Hartree-Fock node of $ ^8S(f^7)$ state. The node has
two topologies: cone times planar surface or a cone "fused"
with planar surface what forms a single sheet surface. There is a smooth transition
between these two forms depending on the positions of six electrons which are
denoted by the small spheres. Note that the node contains the $z-$axis which
is denoted by the dashed line. 
\label{fig:f7}}
\end{figure}

\subsection{\label{sec:level32}  Approximate Hartree-Fock nodes of the $^8S(f^7)$ ion}
We will use similar strategy as in the preceding case. 
After rotating one of the electrons to $z$-axis we
expand the determinant in the probe electron column 
and eliminate the radial orbitals which form
an overall prefactor of the Slater determinant since we assume that
all seven $f$-states are from the same $l=3$ subshell (eg, 4$f$).
We get 
\begin{equation}
(m_1x+m_2y)(4z^2-x^2-y^2)+m_3z(x^2-y^2)+m_4xyz\nonumber
\end{equation}
\begin{equation}
+m_5x(x^2-3y^2)+m_6y(y^2-3x^2)=0
\end{equation}
Note that the node  contains the $z$-axis 
and there are $two$ possible values of $z$ for any $x,y$ since 
the form is quadratic in $z$. This restricts the node 
shapes significantly and by further analysis one can find that
the nodal surface projection into 3D has two topologies 
(Fig. \ref{fig:f7}). The first one
is  a cone times a planar surface (topologically
equivalent to the $Y_{40}$ spherical harmonic).
Note that, in general, the  planar surface is deformed from
a straight plane since it passes
through the origin and, in addition, it fits three of the electrons.
The second topology is 
a "fused" cone and planar surface which results in a general single
sheet cubic surface.
The node transforms smoothly between 
these two topologies depending on how the six 
electrons move in space.
These two topologies define the projection of the node 
into the probe 3D space and therefore enable us to capture the 
many-dimensional node for this particular Hartree-Fock state. 
This again enables to describe the complete node using
a theorem from algebraic geometry which states that any cubic 
surface is determined by an appropriate 
mapping of six points in a projective
plane \cite{pedoe,semple,harris}. To use it we first need to
realize the following property of  
the 3D projected node: The node equation above contains
only a homogeneous polynomial in $x,y,z$
which implies that in spherical coordinates
the radius can be eliminated and the node is dependent 
only on angular variables.
Hence, any line defined by an arbitrary point on the node
and the origin (ie, a ray) lies on the node.
In other words, 
we see that the surface is ruled, ie, it can be created by continuous 
sweep(s) of ray(s)
passing through origin. This enables us to project the positions of the six
electrons on an arbitrary plane which does not contain the origin and 
the node will cut such a plane in a cubic curve. As we mentioned above,
a theorem from
the algebraic geometry of cubic surfaces and curves says 
that any cubic surface is fully described by 
six points 
in a projective plane (see \cite{pedoe,semple,harris}). 
For ruled surfaces any plane not passing
through the origin is a projective plane and therefore we can specify
a one to one correspondence
between the $6\times 3=18$ dimensional space and our cubic surface
in 3D. Obviously, there will be a number of lower-dimensional 
nodes which will correspond to positions
of electrons with additional constraints such as when they lie on 
curve with the degree lower than cubic; ie, a conic.

\begin{figure}[ht]
\centering
\begin{tabular}{cc}
{\resizebox{3.4in}{!}{\includegraphics{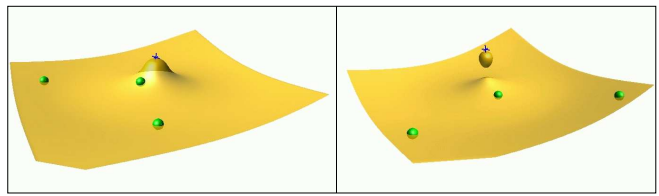}}}
\end{tabular}
\caption{(Color online) 
The 3D projection of the nitrogen cation $^5S(sp^3)$
Hartree-Fock node (the core electrons are eliminated by pseudopotentials). 
The projected node exhibits two
topologies. It is either
a planar surface deformed by the radial orbital functions 
at the nucleus or,
in certain configurations, the deformation forms a small bubble
detached form the surface (the picture on the right). 
The small cross is the location of the ion while the small spheres
denote positions of electrons.
\label{fig:sp3}}
\end{figure}

\subsection{\label{sec:level33}Approximate Hartree-Fock nodes of the $^5S(sp^3)$ ion}
The HF node for  this two-shell spin-polarized 
state can be investigated in a similar way as in previous cases
with a new feature that the radial parts will
be present in the expansion of the determinant.
By expanding the determinant in the column
of the probe electron with position $x,y,z$ 
the 3D node projection is simply given by
\begin{equation}
x+b'y+c'z+ d'\eta(r)=0
\end{equation}
where $b',c',d'$ depend on ratios of cofactors 
and $\eta(r)=\rho_s(r)/\rho_p(r)$
is the ratio of radial parts of $s$ and $p$ orbitals 
and $r=\sqrt{x^2+y^2+z^2}$.
The probe electron will see a plane with a 
approximately bell-shape deformation in the area of the nucleus 
(See Fig. \ref{fig:sp3}).
The shape of deformation depends on the ratio of $s$ and $p$ radial parts and the 
magnitudes and signs of the cofactors.
For certain 
configurations the deformation is so large that it gets detached
from the surface and forms a separated ellipsoid-like bubble.
The bubble results from the radial dependence of $\eta(r)$ which  
for pseudized core is not a monotonic function and therefore
can create new topologies.
Note that despite the fact that the 3D projection shows a separated
region of space (the bubble) the complete node has again 
the minimal number of nodal cells
property. To understand this, suppose that the probe electron is
located inside the bubble and wavefunction there has a positive sign.
Let us try to imagine how the electron can get
to the other positive region (the other side of
the planar surface).
Seemingly, the electron would need to cross the nodal
surface twice (the surface of the bubble and the planar surface).
However, the complete node is a collective-coordinate object 
and by moving the other two electrons
in an appropriate way the bubble attaches to the surface and
then fuses into a single surface (Fig. 5, left)
so that the probe electron can reach 
the positive region without node crossing.

In order to see whether the correlation would change the HF node
we have carried out a limited study of the CI wavefunction nodes for this
case; we have found some differences but we have not discovered any
dramatic changes 
to the HF nodes.  To quantify 
this further we have calculated the
CI energy (with the same basis and level of correlation as in the previous 
cases) and the result is in the last row of Tab. \ref{tab_xx}. We estimate that 
the fixed-node bias of the HF node
is of the order of   $\approx$ 0.001 Hartree which is close to $\approx$
 3 \% of 
the correlation energy. Obviously,  the DMC energy is above
the exact one and percentage-wise the amount of missing 
correlation energy is
not insignificant. We conjecture that
the HF node is reasonably close to the exact one although the 
fine details of the nodal surface are not captured perfectly.
\subsection{\label{sec:level34} Approximate Hartree-Fock nodes of 
spin-polarized $p^3d^5$  and $sp^3d^5$ shells with $S$ symmetry}
\begin{figure}[ht]
\centering
{\resizebox{3.4in}{!}
{\includegraphics{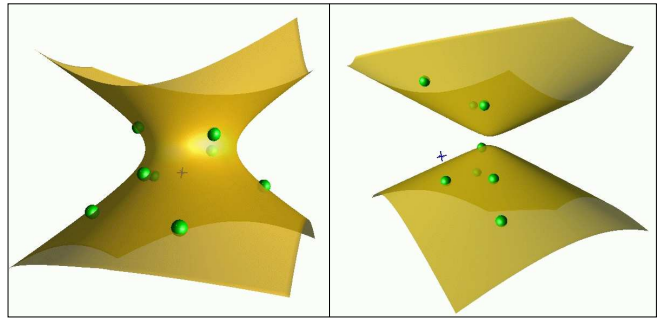}}}
\caption{(Color online) 
The 3D projection of the angular part of the $^{10}S(sp^3d^5)$ state
 Hartree-Fock node 
(with radial parts of orbitals identical for all $spd$ orbitals).  
The projection has a topology
of a single-sheet or double-sheet hyperboloid. The small cross shows the 
location of the nucleus while the spheres illustrate the electron positions. 
\label{fig:sp3d5}}
\end{figure}

\begin{figure}[ht]
\centering
{\resizebox{3.4in}{!}
{\includegraphics{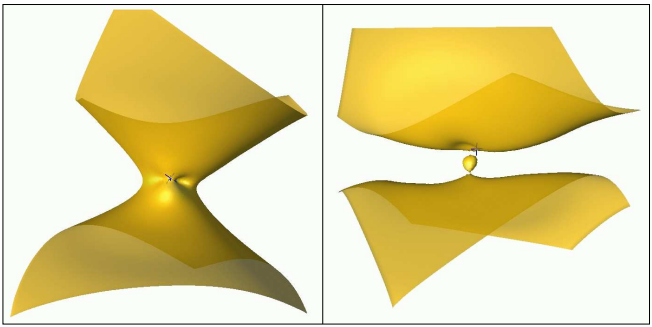}}}
\caption{ (Color online) Projected Hartree-Fock node of
$^{10}S(sp^3d^5)$ of the majority spin valence electrons in
$\mathrm{Mn}^{+2}$ ion. The Ne-core electrons are eliminated 
by pseudopotentials. Note the deformations from the radial
parts of orbitals, including a  
small bubble detached from the rest of the surface (the right picture). 
For clarity, 
the positions of electrons 
have been omitted.   
\label{fig:Mn}}
\end{figure}

Let us for a moment assume a model wavefunction
 in which the radial parts of $s,p,d$
orbitals are identical. Then, using the arrangements similar
to $d^5$ case, we can 
 expand the determinant of $p^3d^5$ in one column and 
for the 3D node projection we then get
\begin{equation}
2u^2-v^2 -w^2 +\alpha u +\beta v +\gamma w=0
\end{equation}
where $u,v,w$ are appropriate linear combinations of $x,y,z$. 
This can be further
rewritten as 
\begin{equation}
2(u+\alpha/4)^2-(v-\beta/2)^2 -(w-\gamma/2)^2 +\delta_0 =0
\end{equation}
where
$\delta_0=(-\alpha^2/2+\beta^2+\gamma^2)/4$.
It is clear that the 
 quadratic surface is offset
from the origin (nucleus)
by a vector normal to 
$\alpha u +\beta v +\gamma w =0$
plane.
Using the properties of quadratic surfaces one finds that
 for $(\alpha^2/(\alpha^2+\beta^2+\gamma^2))<2/3$ the node
is
a single-sheet hyperboloid with the radius $\sqrt{\delta_0}$; otherwise
it has a shape of
 a double-sheet hyperboloid. The double-sheet hyperboloid forms when
there is 
an electron located close to the origin.
A special case is a cone which corresponds to ($ \delta_0=0$).
The case of $sp^3d^5$ is similar, but with different $\delta_0$
which now has a contribution from the $s$-orbital (see Fig. \ref{fig:sp3d5}). 
Once we include also the correct radial parts of orbitals in the 
$s,p,d$ channels 
the coefficients of the quadratic form
depend on both cofactors and orbital radial functions.
The resulting
 nodal surface is deformed beyond an ideal quadric and shows some
more complicated structure around the nucleus (see Fig. \ref{fig:Mn})
as illustrated on HF nodes of the majority spin electrons in Mn$^{++}$ ion
(note that the Ne-core electrons were eliminated by pseudopotentials).

\section{\label{sec:level4} Generalized electron coordinates}
What we have learned from the previous cases is that
for small number of electrons
the Hartree-Fock wavefunctions 
display nodes which,  if transformed in an appropriate way,
 lead to rather simple geometries.

In addition, it is instructive to consider
how the nodal surfaces evolve with
increasing number of electrons. Obviously,
HF theory leads to low kinetic energies and the resulting 
mean-field nodes are very smooth. 
The exact nodes of the high symmetry $P(p^2)$ and $S(p^3)$ cases
can be interpreted as   
{\em reoriented} planes which enable us to fit   
 one or two electrons and obviously such
 rotations do not cause any increase in the kinetic energy.
For more particles the rotations and translations are not
sufficient to fit the electron positions 
and the lowest increase in kinetic energy in atomic systems
is apparently produced by rescaling of the axes.  

Finally, for larger number of 
electrons the node becomes more deformed and complex
with possibilities of new topologies and topological
changes. 
 The 3D node projections we have analyzed above show 
that often there exist
coordinate transformations which can
simplify the node description and enable us to find useful
node parametrizations (at least,
for our cases of spin-polarized electronic subshells in ions). 
As we demonstrated on the
$^3P(p^2)$ state similar spin-polarized open shells 
can be studied by analogous techniques as well.

In the analysis of the nodes we have presented a number of
coordinate transformations and maps which enabled us to understand
particular nodal structures and their topologies. It is interesting
to explore this idea further and think about possible research
directions. 
In order to illustrate some of the possibilities
let us define
a single-electron 
coordinate ${\bf r}_i \to {\bf r}^*_i$ transformation 
as
\begin{equation}
{\bf r}^*_i = {\bf M}_i(R){\bf r}_i + {\bf t}_i(R)
\end{equation}
where
${\bf M}_i(R)$ is a metric tensor, 
${\bf t}_i(R)$ is a spatial offset (translation)
while $R$ represents coordinates 
of all electrons. 
The dependence of ${\bf M}_i(R)$ and ${\bf t}_i(R)$ on $R$
can be nonlinear and even include 
an explicit  dependence on external potential to
describe charge
inhomogeneities or required symmetries.
The antisymmetry condition furher restricts the dependences of 
${\bf M}_i(R)$ and ${\bf t}_i(R)$ on $R$. Obviously
the metric tensor has to be positive definite but otherwise the
transformation is variationally free.
Actually, 
the usefulness of this transformation can be twofold.
 First, as we mentioned, it can be employed
as an efficient way to project the high-dimensional 
nodal manifold
into a simpler, low-dimensional projection. Second, it can
be useful as a new 
variational freedom to optimize the wavefunction nodes using  
the following form of a single determinant or linear combination
of determinants
\begin{equation}
\Psi_{}=\sum_kd_k{\rm det}_k\{\varphi_{\alpha}^{(k)}({\bf r}^*_i)\}
\end{equation}
where $\{\varphi_{\alpha}^{(k)}\}$ are one-particle orbitals
and $d_k$ are expansion coefficients.
Therefore, besides optimizing the orbitals, one can also
optimize the metric tensor and offset
in order to get better variational wavefunctions and fixed-node energies.
In fact,
the transformation above can be considered a generalization of
 the Feynman-Cohen backflow quasi-particle 
coordinates \cite{feynman}.
By simplifying
${\bf M}_i(R)$ to the unit matrix times a scalar function
 we can easily recover
 the backflow wavefunction which Feynman and Cohen suggested
for liquid helium 
\cite{feynman} and which was successfully
employed in QMC in several previous studies \cite{kwon,markus,panoff}. 
The new 
feature proposed here
is the metric tensor which enables to better describe the systems with
 inhomogeneities and/or with rotation symmetries.

\section{\label{sec:level5} Conclusions}
We have investigated
the nodes of atomic and molecular spin-polarized systems
with one-particle states in $s,p,d$ channels.
We have studied cases with high symmetries which 
enabled us to find exact nodes for several 
 states  with a
few electrons ($p^2, p^3, \pi^2$). 
Moreover, the projection of multi-dimensional 
manifolds into 3D space enabled us to 
study and characterize properties of nodes, in particular,
their topologies 
for the Hartree-Fock wavefunctions. 
This analysis has provided useful insights and 
enabled us to formulate
a general transformation of one-particle coordinates 
using coordinate translation (backflow) and metric tensor
to capture inhomogeneities and/or rotation symmetries.
Such transformations can be useful for understanding the nodal
properties and topologies and also as a new variational
freedom for QMC trial wavefunctions.

\begin{acknowledgments}
The support by ONR-N00014- 01-1-0408,
and NSF DMR-0121361, DMR-0121361 grants 
is gratefully acknowledged. L.M. would like to thank
D. Bressanini and P. Reynolds for discussions and N. 
Salwen for help in the early stages of this work.
\end{acknowledgments}

\end{document}